\documentclass[fleqn,usenatbib]{mnras}

\usepackage{hyperref}

\usepackage[T1]{fontenc}
\usepackage{ae,aecompl}

\usepackage{newtxtext,newtxmath}
\usepackage{graphicx}
\usepackage{color}
\usepackage{soul}
\usepackage{url}
\usepackage{bm}         
\usepackage{times}
\usepackage{dcolumn}
\usepackage{bm}
\usepackage{epsf}
\usepackage{amssymb}
\usepackage{caption}
\usepackage{subcaption}
\newcommand{\be}{\begin{equation}}
\newcommand{\ee}{\end{equation}}

\newcommand{\beq}{\begin{equation}}
\newcommand{\eeq}{\end{equation}}
\newcommand{\beqn}{\begin{eqnarray}}
\newcommand{\eeqn}{\end{eqnarray}}

\usepackage{color}

\title[Fall-back with r-process heating]{Imprints of r-process heating on fall-back accretion: distinguishing black hole-neutron star from double neutron star mergers}

\author[D.~Desai et al.]{
D.~Desai$^{1}$, B.~D.~Metzger$^{1}$, F.~Foucart$^{2}$\\
$^{1}$ Department of Physics and Columbia Astrophysics Laboratory, Columbia University, New York, NY 10027, USA\\
$^{2}$ Department of Physics \& Astronomy, University of New Hampshire, 9 Library Way, Durham NH 03824, USA\\
}

\begin{document}

\label{firstpage}
\pagerange{\pageref{firstpage}--\pageref{lastpage}}
\maketitle

\begin{abstract}
Mergers of compact binaries containing two neutron stars (NS-NS), or a neutron star and a stellar-mass black hole (NS-BH), are likely progenitors of short-duration gamma ray bursts (SGRBs).  A fraction $\gtrsim 20\%$ of SGRBs are followed by temporally-extended ($\gtrsim$ minute-long), variable X-ray emission, attributed to ongoing activity of the central engine.  One source of late-time engine activity is fall-back accretion of bound tidal ejecta; however, observed extended emission light curves do not track the naively-anticipated, uninterrupted $t^{-5/3}$ power-law decay, instead showing a lull or gap in emission typically lasting tens of seconds after the burst.  Here, we re-examine the impact of heating due to rapid neutron capture ($r$-process) nucleosynthesis on the rate of the fall-back accretion, using ejecta properties extracted from numerical relativity simulations of NS-BH mergers.  Depending on the electron fraction of the ejecta and the mass of the remnant black hole, $r$-process heating can imprint a range of fall-back behavior, ranging from temporal lulls of up to tens of seconds to complete late-time cut-off in the accretion rate.  This behavior is robust to realistic variations in the nuclear heating experienced by different parts of the ejecta.  Central black holes with masses $\lesssim 3M_{\odot}$ typically experience absolute cut-offs in the fall-back rate, while more massive $\gtrsim 6-8M_{\odot}$ black holes instead show temporal gaps.  We thus propose that SGRBs showing extended X-ray emission arise from NS-BH, rather than NS-NS, mergers.  Our model implies a NS-BH  merger detection rate by LIGO which, in steady-state, is comparable to or greater than that of NS-NS mergers. 
\end{abstract}


\section{Introduction}

Short-duration gamma ray bursts (SGRBs) are commonly believed to be powered by rapid accretion onto a rapidly-spinning black hole, following the coalescence of a compact binary system (e.g.~\citealt{Narayan+92}).  The latter may be comprised either of two neutron stars (NS-NS) or a neutron star and a stellar-mass black hole (NS-BH).  The recent discovery of an SGRB \citep{LIGO+17Fermi} coincident with the gravitational wave event GW170817 \citep{LIGO+17DISCOVERY}, as well as non-thermal emission from the off-axis afterglow of a relativistic jet (e.g.~\citealt{Hallinan+17,Alexander+17,Margutti+17,Troja+17,Haggard+17}), provided compelling evidence that at least some SGRBs arise from NS-NS mergers \citep{Blinnikov+84,Paczynski86,Goodman86,Eichler+89}.  Although no NS-BH binaries are currently known, they are theoretically predicted to exist and may also give rise to SGRB emission in cases where the merger results in the creation of an accretion disk (e.g.~\citealt{Rosswog05,Kyutoku+11,Foucart+13,Foucart+14,Foucart+16,Paschalidis+15,Bhattacharya+18}).  This latter condition requires that the BH be of sufficiently low mass and/or rapidly spinning in the prograde direction with respect to the binary orbit, such that the NS will be tidally disrupted before falling into the BH horizon (e.g.~\citealt{Foucart12}).

\begin{figure}
  \begin{center}
  \includegraphics*[width=.5\textwidth]{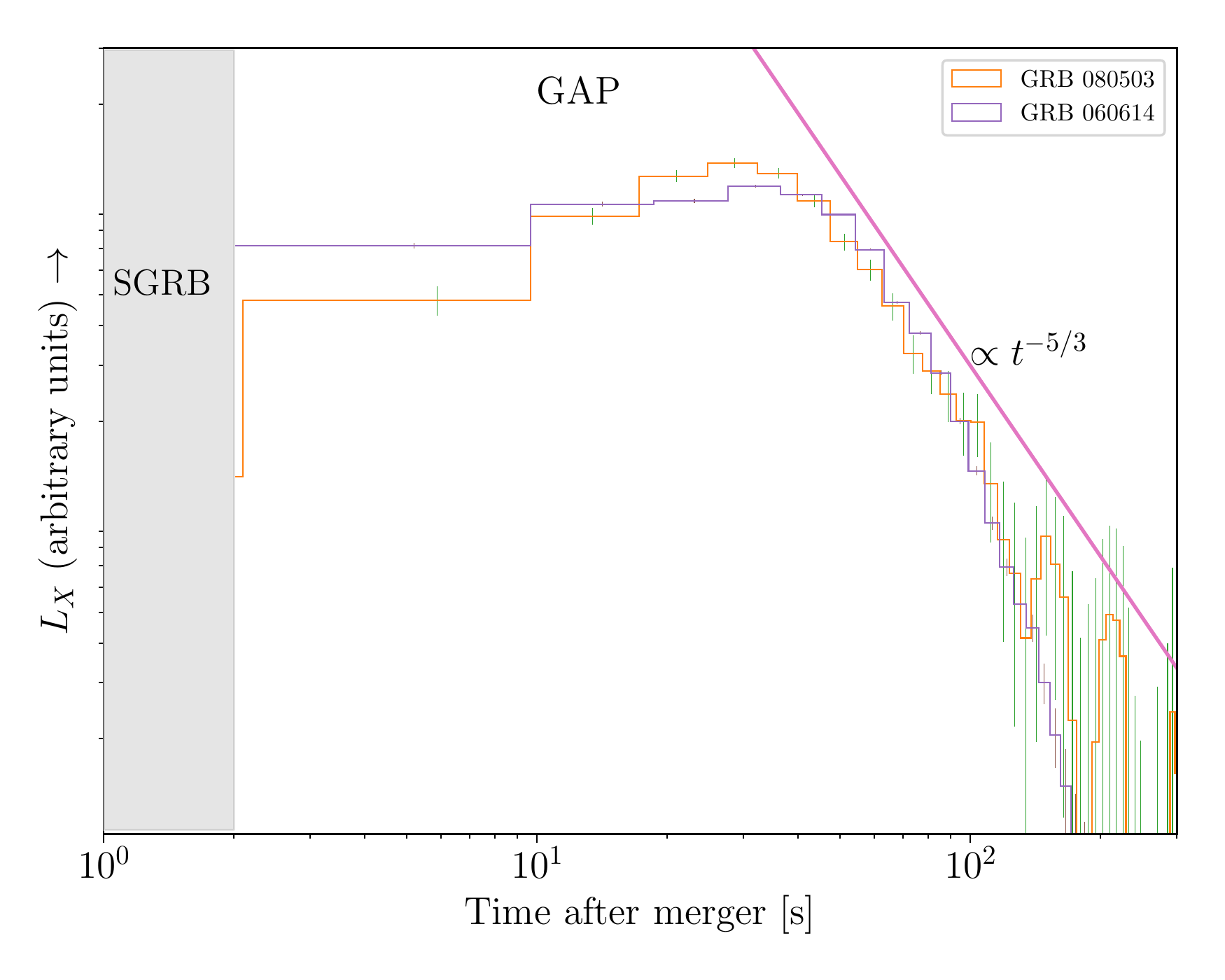}
  \caption{{\it Swift} BAT X-ray light curves of two SGRBs, GRB080503 \citep{Perley+09} and GRB 060614 \citep{Gehrels+06}, which show their temporally-extended prompt emission.  Both cases show remarkably similar light curves, which peak on a timescale $\sim 30$ s after the prompt SGRB spike of duration $\lesssim 2$ s (gray region; not shown at the chosen time binning).  The lull in emission from $t \sim 2-30$ s contrasts with the naive expectation that the X-ray luminosity track the $\propto t^{-5/3}$ mass fall-back accretion (shown for comparison with a pink line).}
  \label{fig:lc_grbee}
  \end{center}
\end{figure}

At least $\sim$ 20\% of SGRBs are followed by temporally-extended X-ray emission, which lasts $\sim 10-1000$ s or longer after the initial prompt gamma-ray burst (\citealt{Norris&Bonnell06,Gehrels+06,Perley+09,Norris+10,Minaev+10,Kaneko+15,Kisaka+17,Burns+18}); Fig.~\ref{fig:lc_grbee} shows the light curves in two clear cases.  This ``extended emission'' is too luminous and time variable to be synchrotron afterglow emission generated as the GRB jet interacts with the interstellar medium.  Instead, it likely results from ongoing activity (e.g.~prompt energy release in the form of a relativistic jet) from the central compact object remnant left by the merger (however, see \citealt{Eichler+09,Eichler17} for an alternative interpretation).

Several models have been proposed for the engine behind the extended emission.  These include the electromagnetic spin-down of a long-lived millisecond magnetar generated from a NS-NS merger (e.g.~\citealt{Metzger+08,Bucciantini+12,Rowlinson+13,Gao+13,Gompertz+14,Gibson+17}).  While this model remains in contention, the lack of late-time radio detections of short GRBs is beginning to place constraints on the amount of rotational energy released from such magnetar remnants \citep{Metzger&Bower14,Horesh+16,Fong+16}.  A stable magnetar is also disfavored in GW170817 by the relatively weak afterglow \citep{Margutti+18,Pooley+18} and the red colors of the late kilonova emission indicative of black hole formation \citep{Metzger&Fernandez14}; however, no extended prompt emission was observed in this event \citep{LIGO+17Fermi}.

Another possibility is that the extended emission is powered by late-time accretion ("fall-back") onto the black hole of tidal matter which is marginally bound to the system after the merger event \citep{Rosswog07}.  However, a glaring issue with this model are the observed evolution of the extended emission light curves.  The rate of mass fall-back after the merger, which is generally taken as a proxy for engine activity, is usually predicted to follow an uninterrupted power law, $\dot{M} \propto t^{-5/3}$, starting from very early times $\lesssim 0.1$ s after the merger.  This result follows from a distribution of ejecta mass with energy, $dM/dE \propto E^{\alpha}$ that is relatively flat ($\alpha \approx 0$) around $E = 0$ (\citealt{Rees88}).  Fig.~\ref{fig:lc_grbee} shows that a single power-law decay is incompatible with observed extended emission light curves, which generally show a lull and delay until peak of a few to tens of seconds after the initial gamma-ray burst.  Explanations proposed for this behavior include a transition in the jet launching mechanism from neutrino-driven to MHD-powered \citep{Barkov&Pozanenko11}, or differences in the rate of accretion of mass versus magnetic flux \citep{Tchekhovskoy&Giannios15,Kisaka&Ioka15}, as the latter controls the jet power in the Blanford-Znajek process (however, \citealt{Parfrey+15} argue that a jet may be produced even in the absence of net magnetic flux).  

Studies of fall-back accretion in neutron star mergers also generally neglect a robust physical process:
the dynamical influence of radioactive heating by heavy nuclei synthesized by rapid neutron captures ($r$-process) in the decompressing material \citep{Metzger+10a}.  This same heating within the unbound debris is responsible for powering the "kilonova" emission \citep{Li&Paczynski98,Metzger+10} days to weeks after the merger, as was observed following GW170817 (e.g.~\citealt{Coulter+17,Soares-Santos+17,Cowperthwaite+17,Drout+17}).  However, at earlier times, less than a few seconds following mass ejection from the merger, the $r$-process heating rate is orders of magnitude higher.  Energy released by the $r$-process does not qualitatively alter the dynamics of the bulk of the {\it unbound} ejecta \citep{Rosswog+14}. However, it can increase the quantity of ejecta and critically shape the dynamics of the {\it marginally bound} ejecta responsible for fall-back accretion, particularly on the second to minute timescales of relevance to the observed extended emission \citep{Metzger+10a}.  

\begin{figure}
  \begin{center}
  \includegraphics*[width=.5\textwidth]{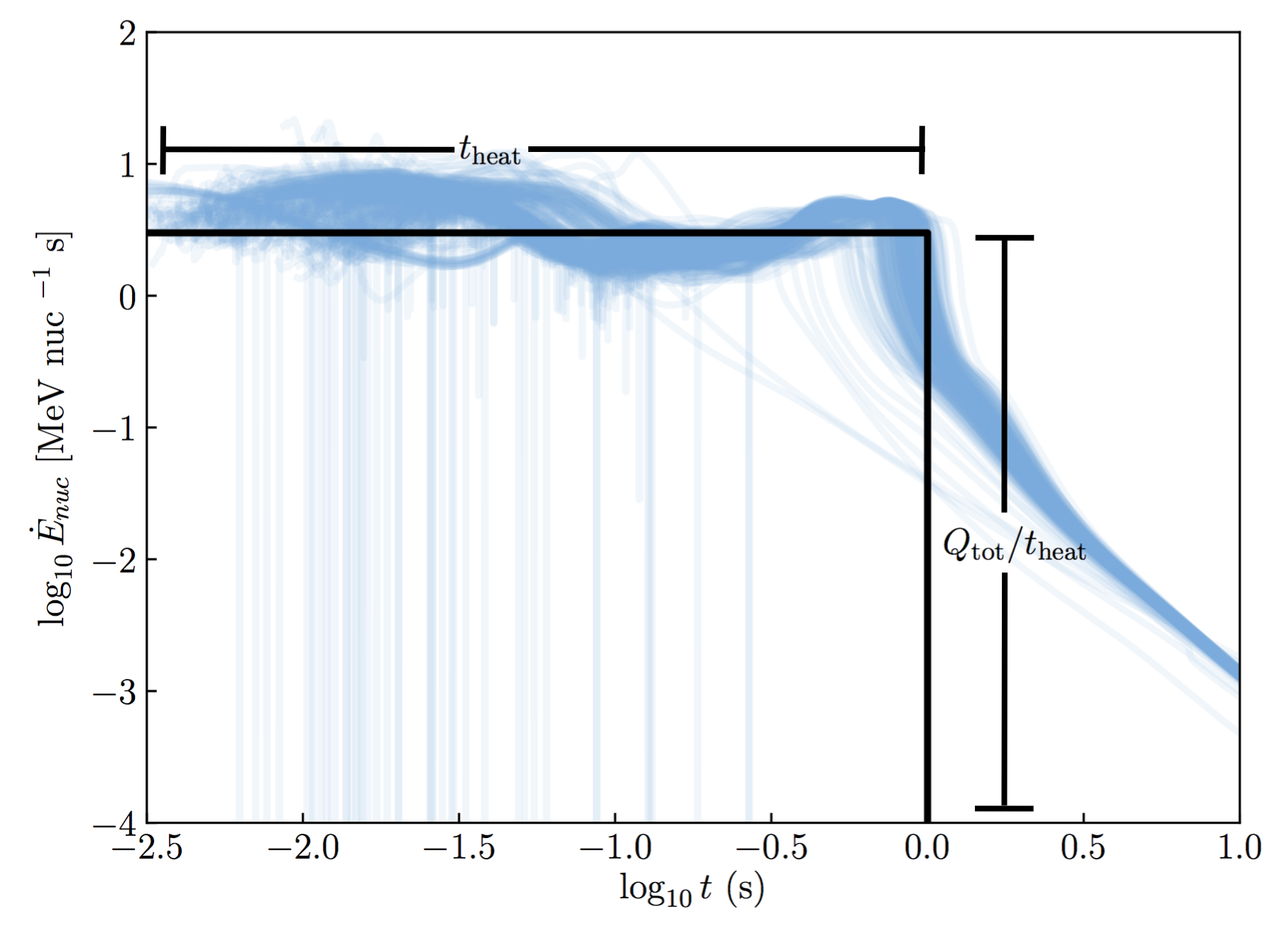}
  \includegraphics[width=0.48\textwidth]{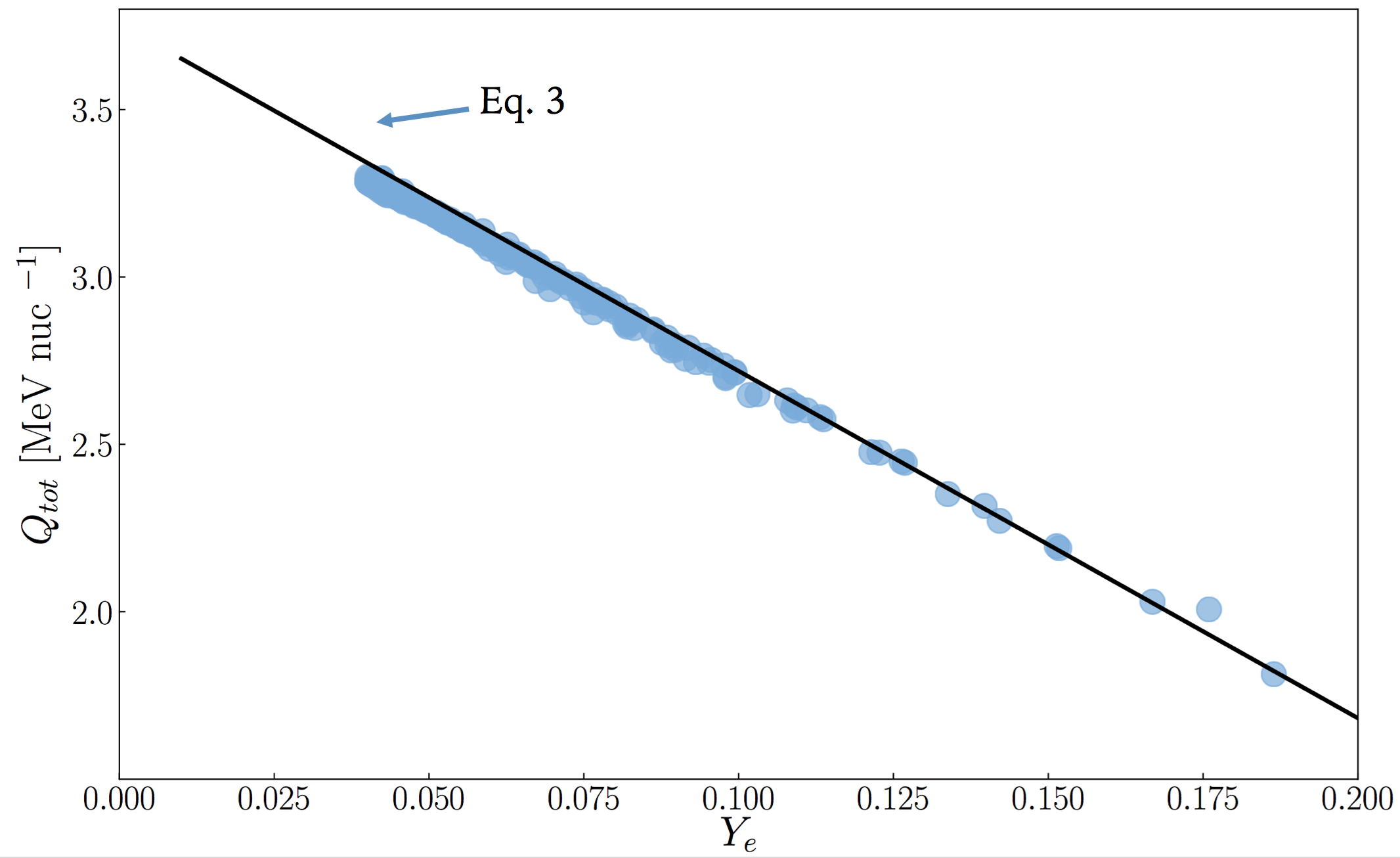}
  \caption{{\bf Top:} Nuclear heating rate of the unbound debris as a function of time since ejection (blue lines), calculated for $\sim$250 separate fluid elements from the NS-BH merger simulation M7\_S8 \citep{Foucart+14} using the SkyNet nuclear reaction network \citep{Lippuner+17}. We have reduced the heating rate from the rate of nuclear energy by a fraction $(1-f_{\nu}) = 0.55$ to account for fraction of the $\beta-$decay energy carried away by neutrinos (eq.~\ref{eq:Qtot}). A black line shows the step-function approximation employed in our fall-back model (eq.~\ref{eq:qdot}), where in this example $Q_{\rm tot}= 3$ MeV and $t_{\rm heat}= 1$ s.  {\bf Bottom:} Total heating rate $Q_{\rm tot}$ for the same fluid elements as a function of their electron fraction $Y_{e}$.  Shown for comparison is the analytic estimate from equation (\ref{eq:Qtot}), for fixed values of the parameters $\bar{A}/\bar{Z} = 2.4$, $f_{\nu} = 0.45$, $\left( \frac{B}{A} \right)_s=8.7$ MeV nuc$^{-1}$ and $\left( \frac{B}{A} \right)_r= 8$ MeV nuc$^{-1}$.  }
  \label{fig:heating_rate}
  \end{center}
\end{figure}

Depending on the electron fraction of the ejecta, $Y_{e} \sim 0.02-0.3$, the $r$-process releases a total energy of $Q_{\rm tot} \sim  1-3$ MeV per nucleon, mostly through beta-decays, at an approximately constant rate over a characteristic heating timescale $t_{\rm heat} \sim 1$ s following ejection (see~Fig.~\ref{fig:heating_rate}).  Then, at times $t \gg t_{\rm heat}$ (e.g. relevant to the kilonova), the heating rate approaches an asymptotic power-law decay $\dot{q} \propto t^{-1.3}$ \citep{Metzger+10}.  A nucleon (mass $m_n$) which is marginally gravitationally bound to the black hole of mass $M$, on an orbit of energy per nucleon $|E_{\rm tot}| = GMm_n/2a$ and semi-major axis $a$ (under the approximation of Newtonian gravity), returns to the hole and circularizes into the accretion disk on a timescale given by its orbital period,
\begin{eqnarray}
t_{\rm orb} = 2\pi\left(\frac{a^{3}}{GM}\right)^{1/2}  
 \approx 1.6\,{\rm s}\left(\frac{|E_{\rm tot}|}{1\,{\rm MeV}}\right)^{-3/2}\left(\frac{M}{5M_{\odot}}\right)
\label{eq:torb}
\end{eqnarray}
This expression reveals several important facts.  First, the total energy available from the $r$-process, $Q_{\rm tot}$, exceeds the binding energy of orbits with fall-back times comparable to observed extended emission after SGRBs ($\gtrsim 10$ s; Fig.~\ref{fig:lc_grbee}); including the effects of $r$-process heating is thus crucial to determining the late-time fall-back rate \citep{Metzger+10a}.  Also of interest is the apparent coincidence that the fall-back time of matter with energy $|E_{\rm tot}| \sim Q_{\rm tot}$ is comparable to the timescale $t_{\rm heat} \sim 1$ s over which the bulk of the heating occurs.  This means that different parts of the debris could receive different amounts of the total available heating, depending on their fall-back time.  \citet{Metzger+10a} show that this can imprint a more complex mass fall-back evolution than the standard $\propto t^{-5/3}$ decay, instead generating either temporal gaps of several seconds or sharp cut-offs in the fall-back rate after a certain time, depending on the ratio of $t_{\rm orb}(|E_{\rm tot}| = Q_{\rm tot})$ and $t_{\rm heat}$.

In this paper we apply the model of \citet{Metzger+10a} in order to estimate the effects of $r$-process heating on the energy distribution of the merger ejecta and its resulting mass fall-back rate, using for the first time initial conditions for the debris properties taken directly from a numerical relativity simulation of a NS-BH merger \citep{Foucart+16}.  We also explore what effects the $Y_e$-dependent nuclear heating rate has, given a realistic spread in the debris properties, on the predicted range of fall-back behavior.  In \S~\ref{sec:model} we describe the problem setup, our treatment of the nuclear heating, and the numerical technique. In \S~\ref{sec:results}, we present our results for the mass fall-back rate.  Finally, in \S~\ref{sec:discussion} we map our findings onto mergers leaving BHs of different mass scales to demonstrate how NS-NS and NS-BH mergers might in principle be distinguished based on the properties of their late-time X-ray light curves.

\section{Model}
\label{sec:model}
\subsection{Numerical Simulation Data}
We use 3D position/velocity data taken from the grid points of the NS-BH merger simulation ``M5-S7-I60'' performed by \citet{Foucart+16} using the numerical code SpEC\footnote{The Spectral Einstein Code: \url{http://www.black-holes.org/SpEC.html}} \citep{Kidder+00}.  This simulation implemented the DD2 Equation of State \citep{Hempel+12} for a $1.4~M_{\odot}$ NS, and included a neutrino leakage scheme, as implemented in \citet{Deaton+13}. The initial mass of the BH is $M_{i} \simeq 5~M_{\odot}$. This is a precessing system, with a dimensionless spin of $\chi = 0.7$ on the BH, prograde but misaligned by $60^\circ$ with respect to the orbital angular momentum.  The final mass of the BH after the merger is $M = 6.11 M_{\odot}$. 

We extract data on the ejecta tidal tail at a time $t =$15 ms after merger.  These are then used as initial conditions in post-processing analysis, in which we evolve the fluid elements further as non-interacting Lagrangian particles.  Because initial energies, positions and velocities used in our analysis were derived from a general relativistic simulation, they are inconsistent with the Newtonian expression for total energy. We therefore rescale the initial velocities from the simulations ($\vec{v}_{\rm GR}$) to their Newtonian equivalent according to $\vec{v}_{\rm N}=\frac{\vec{v}_{\rm GR}}{|\vec{v}_{\rm GR}|}|\vec{v}_{\rm N}|$, where
\beqn
\frac12 m_n{|\vec{v}_{\rm N}|}^2 - \frac{GMm_n}{r} ={} E_{\rm tot},
\label{eq:rescale}
\eeqn
and $E_{\rm tot} = -m_n(u_t+1)$ is an estimate of the specific binding energy (per nucleon) of the test particle, assuming geodesic motion in a time-independent spacetime, and $r$ is the distance from the BH.

\begin{figure}
\begin{center}
\includegraphics*[width=.5\textwidth]{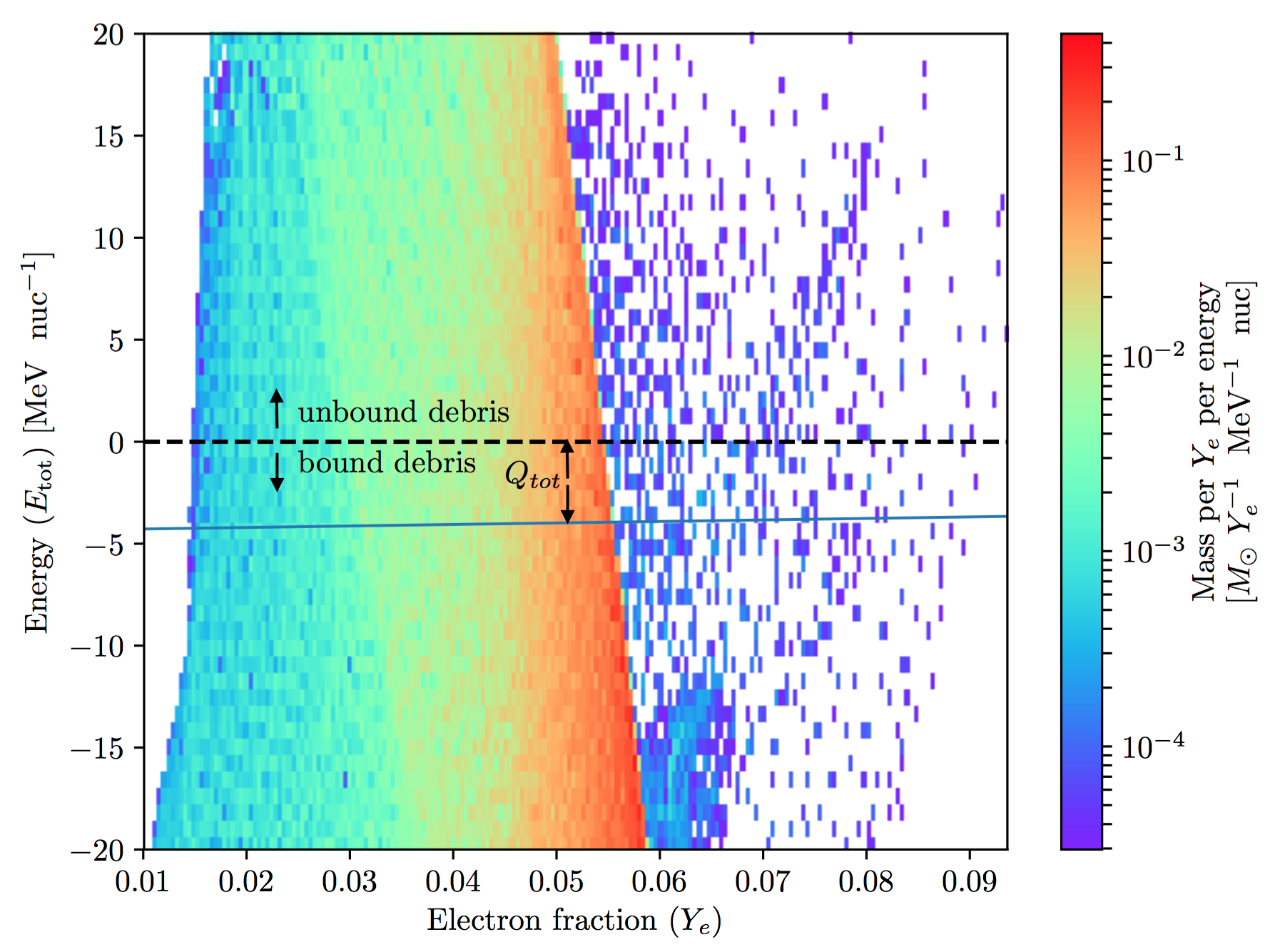}
\caption
{Mass-weighted distribution of the tidal tail ejecta, extracted from the NS-BH merger simulation data at $t = 15$ ms post-merger, in the space of initial energy, $E_{\rm tot}$, and electron fraction, $Y_{e}$.  A dashed line at $E_{\rm tot} = 0$ separates the debris which is initially bound versus unbound to the black hole.  Fluid elements above the solid blue line could be unbound due to $r$-process heating along if the ejecta receives the total available heating, $Q_{\rm tot} \sim 3$ MeV (Fig.~\ref{fig:heating_rate}, bottom panel).}
\label{fig:ye_u}
\end{center}
\end{figure}

Fig.~\ref{fig:ye_u} shows the distribution of initial ejecta properties as a function of $E_{\rm tot}$ and electron fraction $Y_e$, weighted by mass.  Most of the debris is neutron-rich, with an electron fraction $Y_{e} \lesssim 0.05$.  A dashed line separates ejecta which is initially gravitationally bound ($E_{\rm tot} \ge 0$) from unbound material ($E_{\rm tot} \le 0$).  Of the total mass $0.116 M_{\odot}$ of the tidal ejecta, approximately $1.23 \times 10^{-2} M_{\odot}$ is unbound.  Even if they are initially bound to the black hole, fluid elements above the solid blue line could in principle gain sufficient heating from the $r$-process (Fig.~\ref{fig:heating_rate}, bottom panel) during their orbit to become unbound.  As we discuss later, depending on the timescale over which the $r$-process heating is released, matter near this line can also remain bound but fall back to the black hole at a later time than had it experienced zero heating.

\subsection{R-Process Heating}
\label{sec:heating}

We include the effects of $r$-process heating on the trajectories of the fall-back debris in a post-processing step, similar to the model employed by \citet{Metzger+10a}.  The total nuclear energy released as neutrons are captured onto seed nuclei during the $r$-process is approximately given by the difference between the initial and final nuclear binding energies,
\beqn
Q_{\rm tot} \simeq (1-f_{\nu})\left[ \left( \frac{B}{A} \right)_r - X_s \left(\frac{B}{A} \right)_s - X_n \Delta_n \right] \label{eq:Qtot},
\eeqn
minus the fraction $f_{\nu}$ of energy lost to neutrino emission ($f_{\nu} \approx 1/2$ at early times of interest).  Here $X_s = \bar{A} Y_e/\bar{Z}$ is the mass fraction of seed nuclei of average atomic mass number
$\bar{A}$ and charge $\bar{Z}$; $\left( \frac{B}{A} \right)_s=8.7$ MeV nuc$^{-1}$ and $\left( \frac{B}{A} \right)_r= 8$ MeV nuc$^{-1}$ are the average binding energies of seed and final r-process nuclei, respectively; $X_n=1-X_s$ is the neutron mass fraction; and $\Delta_n = (m_n-m_p)c^2= 1.293$ MeV is the neutron-proton mass difference.  For typical values $Y_{e} \approx 0.05$ (Fig.~\ref{fig:ye_u}), $\bar{A} \approx 90$, $\bar{Z} = 38$, $f_{\nu} = 0.45$, we find $Q_{\rm tot} \approx 3$ MeV, consistent with the results of SkyNet nuclear reaction calculations (bottom panel of Fig.~\ref{fig:heating_rate}).

As shown in the top panel Fig.~\ref{fig:heating_rate}, the $r$-process heating in freely-expanding unbound debris is approximately constant for a timescale of $t_{\rm heat} \sim 1$ s, when most of the total energy is released, before rapidly entering a power-law decline.  We approximate this behavior by a function of the simple form, 
\be \label{eq:qdot}
\dot{q} =
\begin{cases}
Q_{\rm tot}/t_{\rm heat}, \hspace{2cm} \text{if\,\,\,}t \le t_{\rm heat} \,{\rm and}\, v_{r} > 0
\\
0,  \hspace{2cm} \text{if\,\,\,}t > t_{\rm heat}  \,{\rm or}\, v_{r} < 0,
\end{cases}
\ee
where $v_{r}$ is the radial velocity of the fluid element and the values of $Q_{\rm tot} \sim 3$ MeV and $t_{\rm heat} \sim 1$ s are parameters which we allow to vary modestly about these fiducial values.

The $r$-process heating, as described by equation (\ref{eq:qdot}), is assumed to terminate if and once matter starts to return to the black hole ($v_{r} < 0$), for reasons we now discuss.  Although the heating rate evolution shown in Fig.~\ref{fig:heating_rate} was calculated for unbound debris, this behavior is a good approximation also for bound debris during its initial outwards motion.  However, the heating is abruptly suppressed once matter reaches apocenter and begins to return to the black hole \citep{Metzger+10a}.  As matter undergoes re-compression, its temperature rises adiabatically and the $r$-process path (which is determined at fixed $Z$ by the equilibrium between neutron capture $(n,\gamma)$ and photodissociation $(\gamma,n)$ processes) is driven closer to the stable valley, where the $\beta-$decay timescales, and thus neutron consumption and energy release timescale, becomes much longer.  \citet{Metzger+10a} show that to good approximation the heating rate effectively shuts off once $v_{r} < 0$, motivating us to neglect heating entirely during re-compression.

\begin{figure}
  \begin{center}

  \includegraphics*[width=.48\textwidth]{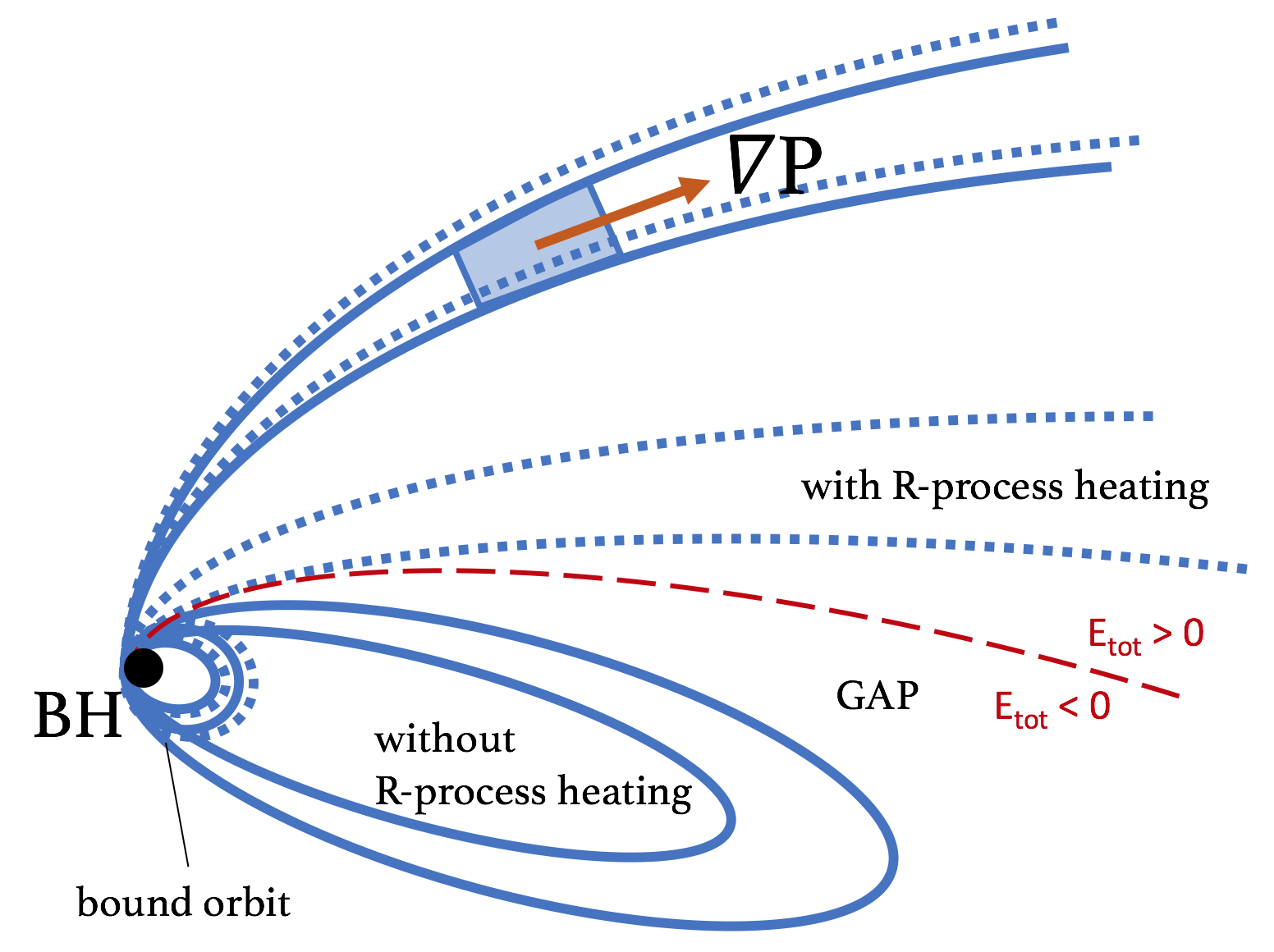}

  \caption
  {Schematic illustration of the tidal ejecta from NS-BH or NS-NS mergers, comparing the fluid element trajectories with (dashed lines) and without (solid lines) the effects of $r$-process heating.  The orbital streams are highly elliptical, such that pressure gradients $\nabla P$ point almost radially outwards and energy released from the $r$-process is transferred quickly (on the expansion timescale) into ejecta kinetic energy.  The trajectories of tightly bound ($E_{\rm tot}\ll 0$) or strongly unbound ($E_{\rm tot}\gg 0$) matter are not greatly altered.  However, marginally bound material with $E_{\rm orb} \gtrsim -Q_{\rm tot}$ and orbital periods comparable to the timescale of the $r$-process experiences preferential heating relative to more tightly bound debris.  This opens a gap in the orbital energy distribution and a temporal gap or late-time cut-off in the mass fall-back rate.}
\label{fig:schematic}
  \end{center}
\end{figure}

\begin{figure}
  \includegraphics[width=0.5\textwidth]{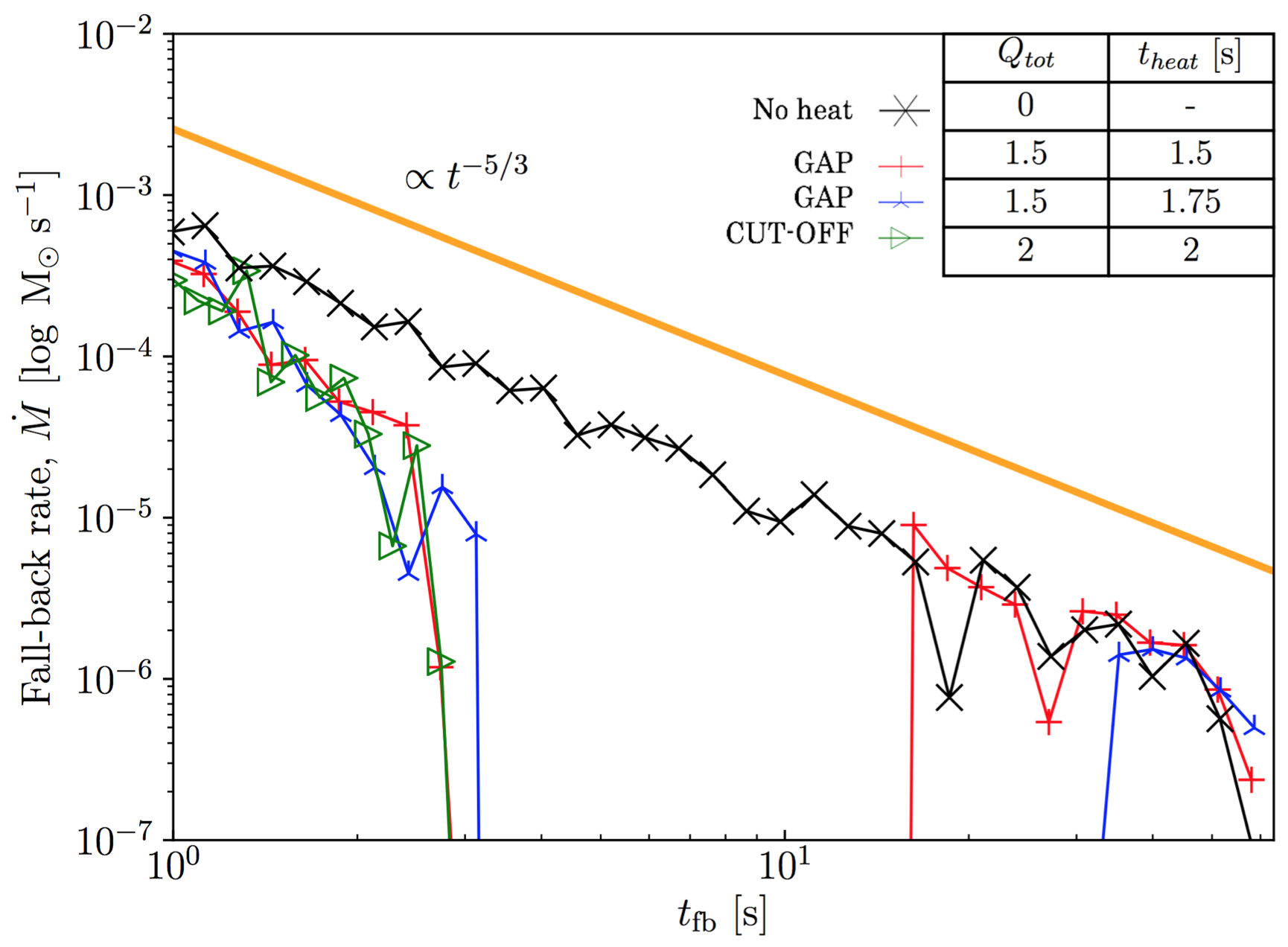}
  \includegraphics[width=0.5\textwidth]{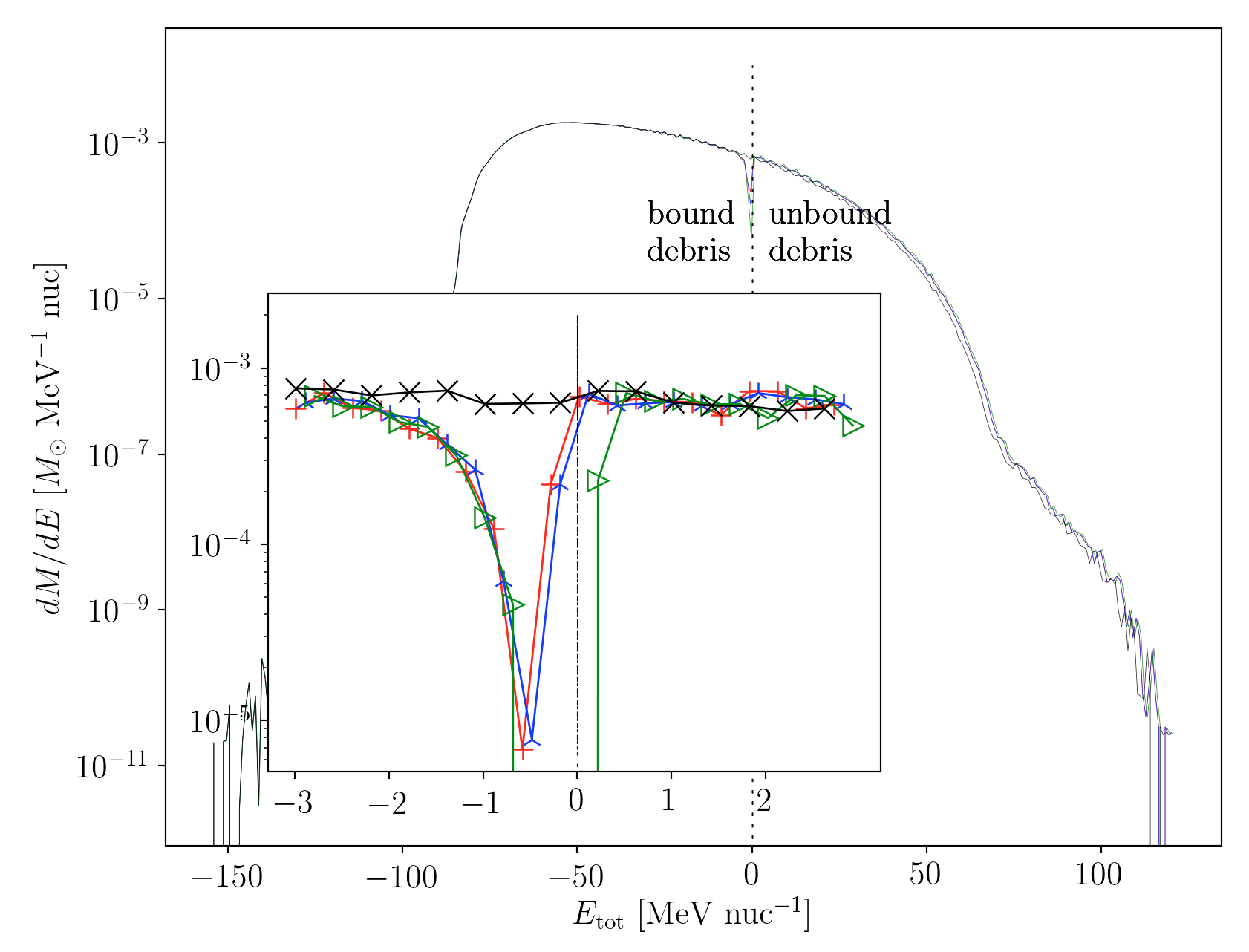}
  \caption{{\bf Top Panel:} Mass fall-back rate, $\dot{M}$, as a function of time, $t_{\rm fb}$, after the NS-BH merger, calculated for different assumptions regarding the total $r$-process heating experienced $Q_{\rm tot}$ and its characteristic duration $t_{\rm heat}$.  Each case is labeled according to the general behavior of the light curve, i.e.~whether heating introduces a temporal gap, or a complete cut-off, in the late-time fall-back rate.  {\bf Bottom Panel:} Distribution of the debris mass with energy, $dM/dE$, comparing the initial distribution after the dynamical phase of the merger to that imprinted by $r$-process heating, shown for the same models and color schemes used in the top panel.  $r$-process heating opens a gap in the energy distribution, which in turn results in either a temporal gap or a complete cut-off in the fall-back rate, depending on whether the energy gap overlaps with $E_{\rm tot} = 0$.}
  \label{fig:fallback}
\end{figure}

\begin{figure*}
  \begin{center}
  \includegraphics[width=1.0\textwidth]{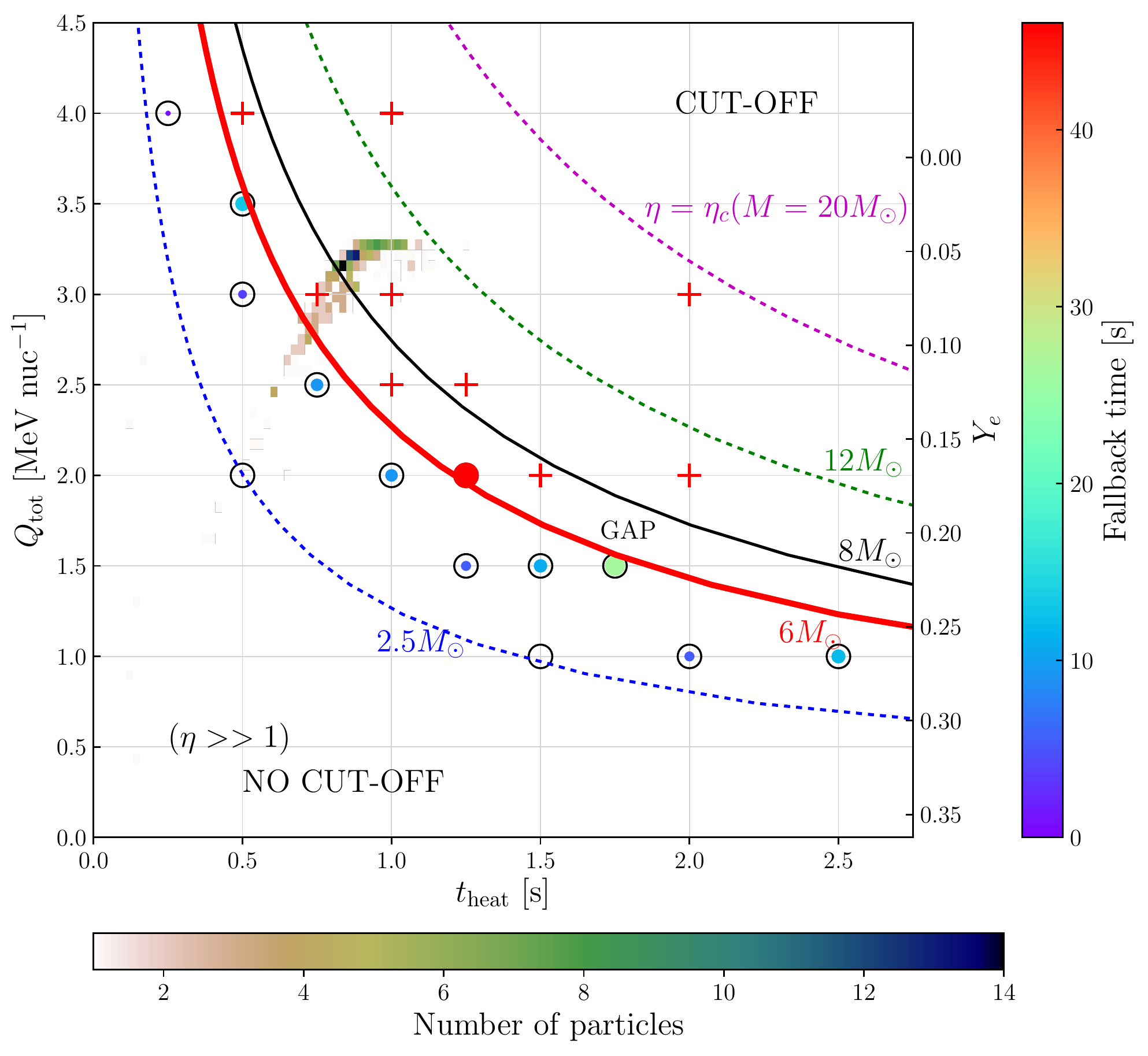}

  \caption
  {Regimes of the impact of $r$-process heating on fall-back accretion in the space of the total nuclear released energy, $Q_{\rm tot}$, and the timescale of the heating, $t_{\rm heat}$.  The value of $Q_{\rm tot}$ (left vertical axis) is mapped onto the initial electron fraction $Y_e$ (right vertical axis) using eq.~(\ref{eq:Qtot}) as shown in the bottom panel of Fig.~\ref{fig:heating_rate}.  A 2D colored histogram shows the mass distribution of the ejecta in ($Q_{\rm tot}, t_{\rm heat}$), calculated by mapping the ejecta properties of our simulation for a NS-BH merger ($M \simeq 6M_{\odot}$) into the $r$-process heating trajectories from SkyNet calculations (Fig.~\ref{fig:heating_rate}).  Symbols show the results of our parameter study in which we assume a BH mass $M \approx 6M_{\odot}$ and that all fluid elements experienced heating characterized by fixed values of $Q_{\rm tot}$ and $t_{\rm heat}$ according to eq.~\ref{eq:qdot} (see Fig.~\ref{fig:fallback} for a few examples).  Crosses denote cases which result in a complete cutoff in the fall-back rate after a given time, while circles show cases in which a temporal gap is opened in the fall-back curve (the duration of the gap is denoted both by the radius and color of the circle using the legend given on the right).  Lines represent the critical condition ($\eta = \eta_{\rm c} \approx 0.95$; eq.~\ref{eq:eta}) giving rise to a long ($\sim 30$ s) gap for different values of the mass of the central black hole as marked, ranging from $M \approx 2.5M_{\odot}$ relevant to NS-NS mergers to $M \sim 6-20M_{\odot}$ relevant to NS-BH mergers.  While the tidal ejecta from NS-NS mergers lies in the cut-off regime ($\eta \gg 1$) for neutron-rich ejecta $Y_e \lesssim 0.2$, the NS-BH merger case resides close to the gap regime ($\eta \sim \eta_{\rm c} \approx 0.95$).}
  \label{fig:param}
  \end{center}
\end{figure*}

\subsection{Numerical Model}

At early times after the merger, the ejecta is dense and highly opaque to photons, such that all of the $r$-process heating (other than that which escapes as neutrinos) goes into internal thermal energy.  The fluid element orbits of interest possess high eccentricities $e$, where
\be 1-e = \frac{r_{\rm p}}{a} \approx 0.03\left(\frac{r_{\rm p}}{5r_{\rm g}}\right)\left(\frac{|E_{\rm tot}|}{3{\rm MeV}}\right). \ee
Here we have normalized the pericenter radius of the debris, $r_p$, to the gravitational radius of the BH, $r_{\rm g} = GM/c^{2}$.  Temperature and density gradients in the ejecta are thus directed nearly radially outwards, such that the $r$-process energy will be transferred through PdV work into ejecta kinetic energy on the local expansion timescale.  Fig.~\ref{fig:schematic} shows a schematic illustration of the influence of $r$-process heating on the trajectories of different fluid elements.

Following explicitly the effects of $r$-process heating on the debris dynamics would require a three-dimensional hydrodynamical simulation across a large dynamical range in radius.  However, given the highly supersonic expansion velocities of the ejecta, such a treatment is not necessary, as the main effect of the heating is a local slow acceleration of the ejecta along the local pressure gradient radial direction and thus the adiabatic conversion of the injected thermal energy to kinetic energy.  Starting with fluid element velocities rescaled from the simulation data (eq.~\ref{eq:rescale}), we directly increase the kinetic energy of the $i$th element according to
\be
\frac{d}{dt}\left(\frac{1}{2}m_n v_{r,i}^{2}\right) = \dot{q}_i,
\ee
where $\dot{q}_i$ follows equation (\ref{eq:qdot}). For each fluid element, we follow its 3D trajectory until either it reaches periapse, which we define as the fall-back time $t_{\rm fb}$, or once the simulation terminates. We then record $t_{\rm fb}$ for all bound material and use the total mass in different bins of $t_{\rm fb}$ to calculate the fall-back accretion rate.


\section{Results for Mass Fall-Back}
\label{sec:results}

The top panel of Figure \ref{fig:fallback} shows examples of the fall-back accretion rate as a function of time after the merger.  In this initial analysis, for each model we assume that all fluid elements experience $r$-process heating rate (according to eq.~\ref{eq:qdot}) characterized by the same total amount $Q_{\rm tot}$ and characteristic duration, $t_{\rm heat}$.  The bottom panel of the figure shows the final energy distribution of the ejecta mass for the same models as compared to the initial energy distribution.

When $r$-process heating is neglected ($Q_{\rm tot} = 0$), the roughly flat energy distribution $dM/dE \approx constant$ around $E_{\rm tot} = 0$, as imparted by the dynamical phase of the merger, is unaltered and the fall-back rate follows the canonical prediction of an uninterrupted $\dot{M} \propto t^{-5/3}$ decay.  By contrast, including the effects of $r$-process heating for values of $Q_{\rm tot} \sim 1-3$ MeV and $t_{\rm heat} \sim $ 1 s drastically changes the energy distribution around $E_{\rm tot} \approx 0$ and results in a more complex fall-back history.  In particular, there are two possible outcomes in shape of the fall-back curve: either an absolute cut-off after some time (e.g. as in the $Q_{\rm tot} = 2$ MeV, $t_{\rm heat}$ = 2 s case), or a cut-off followed by re-emergence of fall-back, i.e.  a "gap" (e.g. as in the $Q_{\rm tot} = 1.5$ MeV and $t_{\rm heat}$ = 1.5 s case). 
As we now discuss, these qualitatively different behaviors can be understood by comparing the timescale over which the ejecta is heated to its orbital timescale (eq.~\ref{eq:torb}; see also \citealt{Metzger+10a}). 

First consider the existence of a {\it critical} orbital period $t_{\rm orb,c}$, which corresponds to matter bound to the BH by an energy equal to the energy $\approx (Q_{\rm tot}/t_{\rm heat})t_{\rm orb}$ it receives from the $r$-process over the orbital period $t_{\rm orb}$ (when $t_{\rm orb} \lesssim t_{\rm heat}$).  Using equation (\ref{eq:torb}) for $t_{\rm orb}$, this gives
\be
t_{\rm orb,c} \approx  0.62\,{\rm s}\left(\frac{Q_{\rm tot}}{3\,{\rm MeV}}\right)^{-3/5}\left(\frac{M}{5M_{\odot}}\right)^{2/5}\left(\frac{t_{\rm heat}}{\rm 1s}\right)^{3/5},
\label{eq:torbc}
\ee
Ejecta which starts on an orbit of period $t_{\rm orb} \gg t_{\rm orb,c}$ always receives the full $r$-process heating before reaching apocenter (at which point the matter starts to re-compress and heating shuts off for reasons discussed earlier), while matter which starts very tightly bound ($t_{\rm orb} \ll t_{\rm orb,c}$) may receive only a fraction $t_{\rm orb}/t_{\rm heat}$ of the total heating (if $t_{\rm heat} \gtrsim t_{\rm orb,c}$).

Crucially, however, if $t_{\rm heat} \gtrsim t_{\rm orb,c}$ even matter with an initial $t_{\rm orb}$ which is slightly less than $t_{\rm orb,c}$ can {\it also receive the full heating} because, as the energy of a fluid element increases, its orbital period also grows, giving it more time to receive the full allotment of nuclear energy (in other words, the final value of $t_{\rm orb}$ diverges in a runaway process due to the $r$-process heating).  This preferential heating opens a gap in the energy distribution of the debris, which can result in an absolute cut-off in the accretion rate, or a temporal gap, depending on the location of the gap.  

Whether such behavior is possible depends on whether the $r$-process heating indeed acts uniformly over the orbit, i.e. on a critical ratio:
\be
\eta \equiv \frac{t_{\rm heat}}{t_{\rm orb,c}} \approx 1.6\left(\frac{M}{5M_{\odot}}\right)^{-2/5}\left(\frac{Q_{\rm tot}}{3\,{\rm MeV}}\right)^{3/5}\left(\frac{t_{\rm heat}}{1\rm s}\right)^{2/5}
\label{eq:eta}
\ee 
If $t_{\rm heat} \ll t_{\rm orb,c}$ ($\eta \gg 1$), then the heating is applied in a short burst uniformly to all fluid elements and there is no runaway (preferential heating) of fluid elements as discussed above.  In this case no significant energy gap is opened; $\dot{M}_{\rm fb}$ shows a slight dip around the time at which $t_{\rm orb} \sim t_{\rm heat} \sim 1$ s, but otherwise experiences no significant interruption of fall-back activity and $\dot{M}$ still approaches a power-law $\propto t^{-5/3}$ decay at later times.

If $t_{\rm heat} \gtrsim t_{\rm orb,c}$ ($\eta \gtrsim 1$) then an energy gap is opened in the debris.
If $t_{\rm heat} \gg t_{\rm orb,c}$ ($\eta \gg 1$), then only the most marginally-bound matter will receive enough heat to unbind before arriving at apocenter and the energy gap extends to $E > 0$.  This case produces an absolute cut-off in $dM/d|E|$ (and hence $\dot{M}_{\rm fb}$) for $|E_{\rm tot}| \lesssim E_{\rm c}$, where $E_{\rm c}$ is the energy of orbits of period $t_{\rm orb} = t_{\rm orb,c}$.


In intermediate cases for which $t_{\rm heat} \sim t_{\rm orb,c}$ ($\eta \sim 1$), a large gap is opened in the energy distribution of the debris, but now material with initial orbital periods $t_{\rm orb} \lesssim t_{\rm orb,c}$ remains marginally bound despite the extra energy it receives, thus opening up a large temporal gap in $\dot{M}_{\rm fb}$ (see Fig.~\ref{fig:schematic} for an illustration).  Specifically, we find that a critical value of $\eta = \eta_{\rm c} \approx 0.95$ is needed to generate a long cutoff of $\sim 30$ s, similar to the observed lulls in the SGRB extended emission light curves (Fig.~\ref{fig:lc_grbee}).  As we discuss later, the black hole mass-dependence of $\eta_{\rm c}$ may have implications for distinct fall-back behavior in NS-BH versus NS-NS mergers.  

Fig.~\ref{fig:param} shows the results of a broader study of the outcomes of fall-back across the parameter space of $r$-process heating parameters $1 \le Q_{\rm tot} \le 4$ MeV and $0 \le t_{\rm heat} \le 3$ s.    Crosses denote runs resulting in cut-off behavior, while circles denote cases with gaps (with the duration of gap indicated by the size of the circle and its color, based on the key given on the right of the diagram).  To highlight the relevant region of parameter space we overlay with a colored histogram the mass-weighted distribution of ($Q_{\rm tot},t_{\rm heat}$), obtained by mapping the ejecta properties from our NS-BH simulation data into the parameters extracted from the SkyNet heating trajectories based on their $Y_e$ values (Fig.~\ref{fig:heating_rate}), where we define $t_{\rm heat}$ as the time at which the heating rate curve first decreases below half of its maximum value.  We also overlay with lines the condition $\eta = \eta_{\rm c}$ (eq.~\ref{eq:eta}) for different assumptions about the BH masses, ranging from low values $M \approx 2.5M_{\odot}$ relevant to the remnants of NS-NS mergers to higher values $M \sim 6-20M_{\odot}$ appropriate to NS-BH mergers.  This location of the crosses and circles relative to the $\eta = \eta_{\rm c}$ line for the BH mass $M \simeq 6 M_{\odot}$ corresponding to our simulation verifies the validity of this criterion as that responsible for separating cut-off from gaps in fall-back behavior.

\subsection{$Y_e$-dependent spread in fluid element heating}

\begin{figure}
  \begin{center}
  \includegraphics[width=0.48\textwidth]{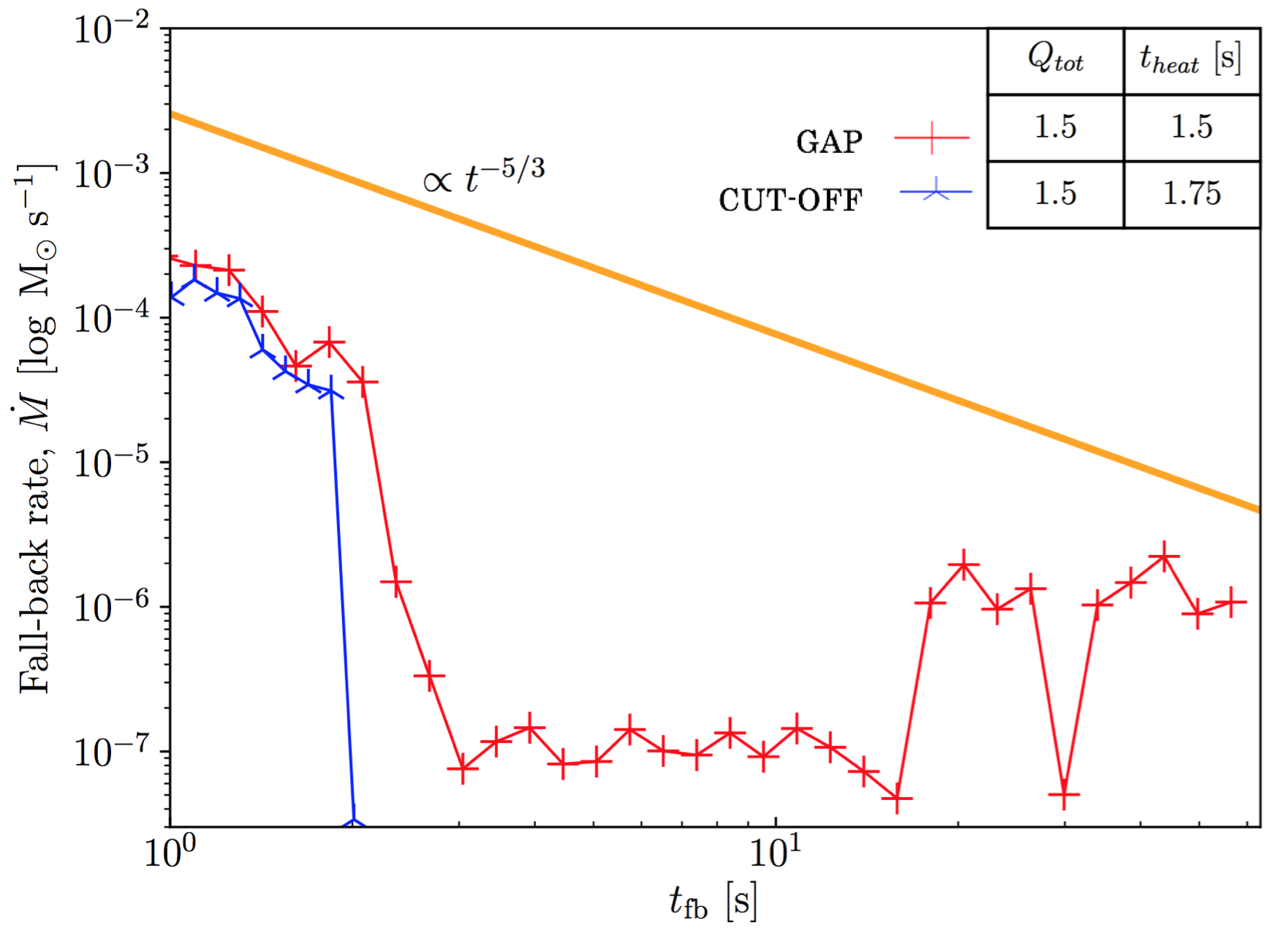}

  \includegraphics[width=0.48\textwidth]{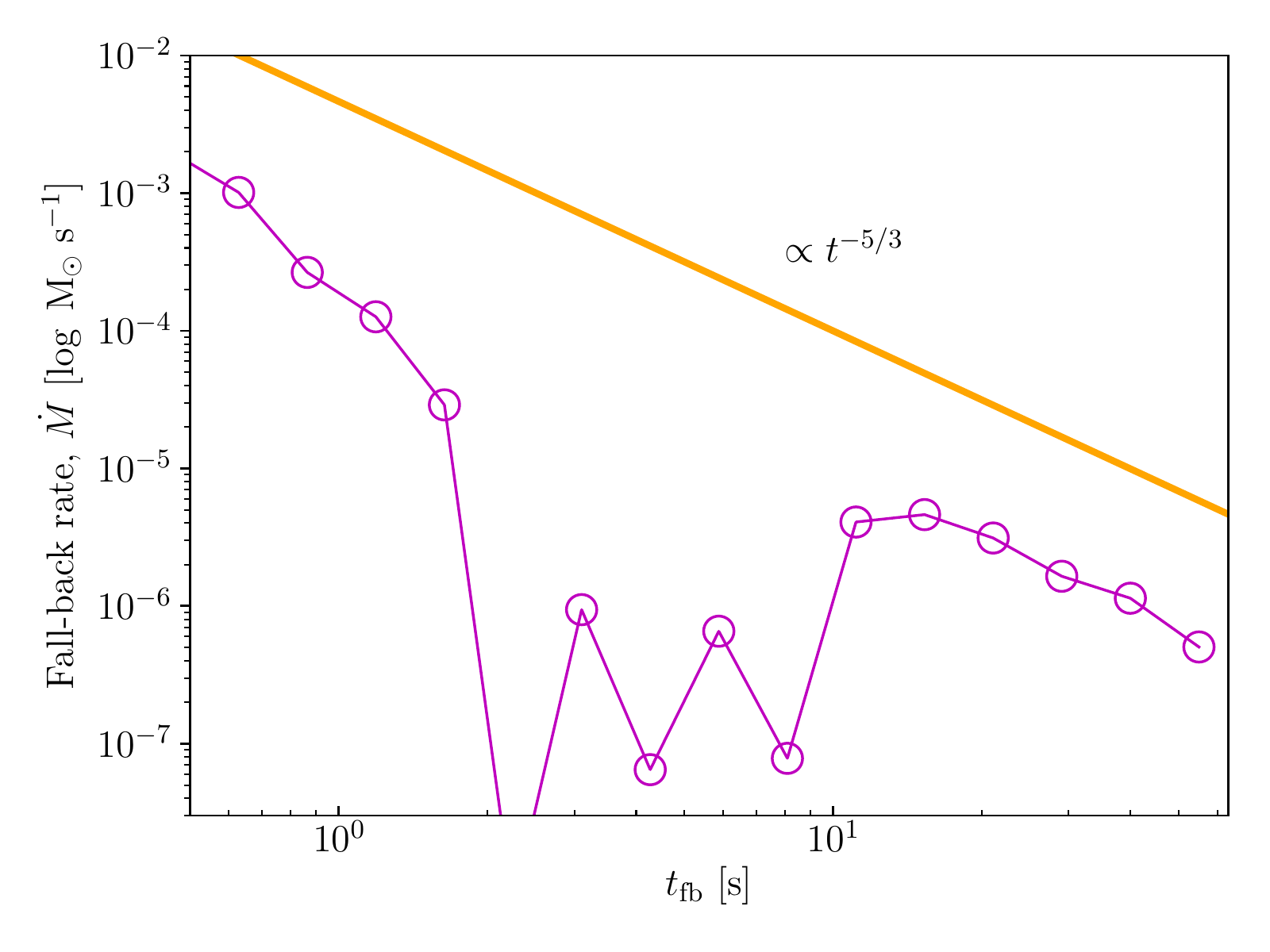}
  \caption{{\bf Top:} Two models from Figure \ref{fig:fallback} showing temporal gaps in the fall-back rate, but now calculated using a realistic spread in the $r$-process heating experienced by different fluid elements due to variations in their $Y_e$ values (eq.~\ref{eq:Qtot}) (see also Fig.~\ref{fig:heating_rate}, bottom panel). The mean values for $Q_{\rm tot}$, around which the spread is centered, are shown in the figure. The previous model with a large temporal gap in the fall-back rate ($t_{\rm heat} = 1.75$ s; {\it blue}) now shows a complete cut-off, while the case with a shorter gap ($t_{\rm heat} = 1.5$ s; {\it red}) has been smoothed out to a lull in accretion.  {\bf Bottom:} Mass fall-back evolution calculated using $r$-process heating curves taken directly from the output of SkyNet simulations (Fig.~\ref{fig:heating_rate}) mapped into our NS-BH simulation data based on their $Y_e$ values.  We see a gap behavior because the ejecta properties lie close to the critical line $\eta = \eta_{\rm c}$ line for the $M = 6M_{\odot}$ black hole (red solid line in Fig.~\ref{fig:param}), but again the gap is partially filled-in due to the spread in heating properties.}
  \label{fig:Yespread}
  \end{center}
\end{figure}

Our calculations shown in Fig.~\ref{fig:fallback} were performed under the assumption that all fluid elements experienced heating characterized by single values of $Q_{\rm tot}$ and $t_{\rm heat}$.  While reasonable as a first-order approximation (Fig.~\ref{fig:heating_rate}), in detail these parameters will vary between fluid elements as a result of a finite spread in their initial electron fraction $Y_{e}$ and precise thermodynamic conditions (e.g. entropy and expansion rate, which affect the properties of the seed nuclei).  It is thus important to address whether the different fall-back outcomes discussed above, particularly the presence of long temporal gaps in accretion, are preserved in the face of such realistic heating variations.

In order to explore the impact of a physical heating spread on our results, we vary the value of $Q_{\rm tot}$ between fluid elements based on their $Y_e$ value (as taken from the simulation data) using equation (\ref{eq:Qtot}; see also Fig.~\ref{fig:heating_rate}, bottom panel).  The other free parameters in the equation (e.g.~$\bar{A}/\bar{Z}$) are chosen to match the {\it mean} values of $Q_{\rm tot}$ and $t_{\rm heat}$ to the comparison cases in which these values are fixed for all fluid elements (e.g., as in the cases shown in Fig.~\ref{fig:fallback})

Figure~\ref{fig:Yespread} shows the effect of $Y_e$-dependent heating on $\dot{M}(t_{\rm fb})$ compared to the two previous models from Figure \ref{fig:fallback} which showed gaps in the fall-back rate.  The case with an initially long gap of $\gtrsim 30$ s was transformed into a complete cut-off by the heating spread.  However, in the case with a shorter gap of $\sim 20$ s the effect of the finite heating spread is to smooth out, but not eliminate, the gap in mass fall-back, in other words turning the "gap" into a "lull".  Our initial conclusion that $r$-process heating can lead to at least partial gaps in the fall-back when the critical condition $\eta \approx \eta_{\rm c}$ is satisfied (where now $\eta$ is defined using the mass averaged values of $Q_{\rm tot}$ and $t_{\rm heat}$) thus appears to be robust.

Our calculations thus far have employed a step-function heating profile as given by eq.~\ref{eq:qdot}.  To explore the sensitivity of our conclusions to this assumption, we perform an identical calculation where we directly using the direct heating curves from SkyNet (Fig.~\ref{fig:heating_rate}, top panel), which have been mapped onto the ejecta from our simulation data according to their closest $Y_e$ value (e.g.~\ref{fig:ye_u}).  We still assume that heating for a given fluid element goes to zero when re-compressing ($v_r < 0$).  The results of this simulation, as shown in the bottom panel of Fig~\ref{fig:Yespread}, is fall-back with a gap of about 10 seconds.  This is because the heating parameters for our fiducial simulation of a NS merging with a $\approx 6M_{\odot}$ BH overlaps with the gap condition $\eta \approx \eta_{\rm c}$ (red solid line in Fig.~\ref{fig:param}).



\hfill \break
\section{Implications for Extended Emission in SGRBs}
\label{sec:discussion}

\begin{figure}
  \begin{center}
  \includegraphics[width=0.48\textwidth]{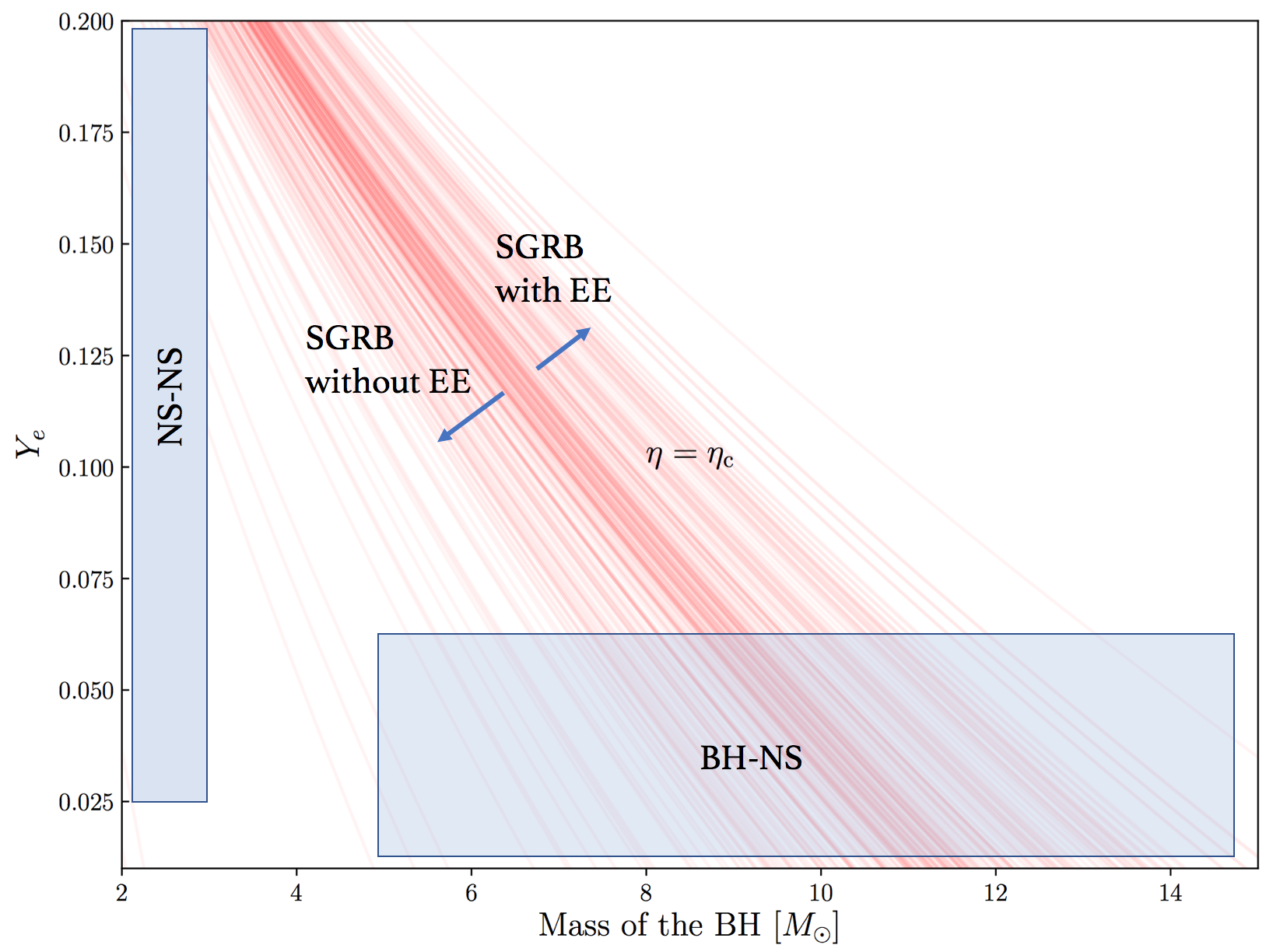}
  \caption{The critical condition $\eta = \eta_{\rm c}$ (eq.~\ref{eq:eta}) in the space of BH mass $M$ and ejecta electron fraction $Y_e$, which separates a temporal gap, versus a cut-off, in the rate of fall-back accretion. The value of $\eta$ is calculated by mapping $Q_{\rm tot}$ to $Y_e$ using equation \ref{eq:Qtot} for different values of $t_{\rm heat}$ as extracted from the results of our SkyNet calculations for the nuclear heating rate. As long as the ejecta is sufficiently neutron-rich ($Y_{e} \lesssim 0.2$), the low-mass black hole remnants of NS-NS mergers are predicted to experience a cut-off in the fall-back rate, while the more massive BHs from NS-BH mergers should experience a gap or lull in fall-back, consistent with the extended emission observed after a fraction of SGRBs (Fig.~\ref{fig:lc_grbee}).}
  \label{fig:YeM}
  \end{center}
\end{figure}

As described in $\S$\ref{sec:results}, the condition $\eta = \eta_{\rm c} \approx 0.95$ separates two distinct regions in the space of $Q_{\rm tot}-t_{\rm heat}$ shown in Fig.~\ref{fig:param}: the lower left corner  ($\eta \ll \eta_{\rm c}$), where fall-back has a gap or is uninterrupted, and the upper right ($\eta \gg \eta_{\rm c}$), where fall-back exhibits a complete cutoff.   

The heating properties of the ejecta from our NS-BH simulation, which left a black hole of mass $M \simeq 6M_{\odot}$, lies in the gap region close to the solid red line.  We have confirmed this behavior by calculating the fall-back rate directly using SkyNet heating trajectories (Fig.~\ref{fig:Yespread}, bottom panel).  However, because $\eta \propto M^{-2/5}$ (eq.~\ref{eq:eta}), otherwise similar ejecta from a merger that resulted in central black hole of lower or greater mass would instead result in $\eta \gg \eta_{\rm c}$ or $\eta \ll \eta_{\rm c}$, respectively, and thus would exhibit qualitatively different fall-back behavior.   Figure \ref{fig:YeM} shows the condition $\eta = \eta_{\rm c}$, now in the space of black hole mass $M$ and electron fraction $Y_e$, where we have used the relationship $Q_{\rm tot}(Y_e)$ from eq.~(\ref{eq:Qtot}) (see bottom panel of  Fig.~\ref{fig:heating_rate}).

If the X-ray luminosity of the extended prompt emission following a short GRB is proportional to the mass fall-back rate, $L_X \propto \dot{M}$, then this reasoning would suggest that NS-NS mergers ($M \approx 2.5M_{\odot}$) with similar ejecta properties would lie in the regime $\eta \gg \eta_{\rm c}$ and thus should generate little or no late-time fall-back and hence would not be accompanied by luminous extended X-ray emission. By contrast, NS-BH mergers, given their more massive BHs $\gtrsim 5M_{\odot}$, lie in the regime $\eta \sim \eta_{\rm c}$ or $\eta \lesssim \eta_{\rm c}$ and thus could produce fall-back with temporal gaps extending up to tens of seconds, in agreement with those short GRBs showing extended emission (Fig.~\ref{fig:lc_grbee}).  

One caveat is that, while these conclusions hold for highly neutron-rich ejecta ($Y_e \lesssim 0.2$), as characterizes the equatorial tidal tails in NS-NS and NS-BH mergers, it would not necessarily apply to the most polar-concentrated shock-heated dynamical ejecta, which experiences stronger weak interactions.  At least in the case of NS-NS mergers, this has been shown to give rise to a wider range of $Y_e$ \citep{Wanajo+14,Goriely+15,Sekiguchi+16}, potentially extending to values $Y_e \gtrsim 0.2$ that would place even NS-NS mergers into the $\eta \sim \eta_{\rm c}$ regime and result in some fall-back.  However, because in many cases the quantity of high $Y_e$ matter is likely to be much less than the total, the amount of fall-back from this component would be significantly smaller and the presence of a cut-off in the fall-back rate might be preserved.

Our results suggest the possibility that the two apparent classes of SGRBs$-$those with and those without extended emission$-$may be associated with NS-BH and NS-NS mergers, respectively.  Such a dichotomy of origin was previously suggested on the completely different basis of the observed distribution of spatial offsets of short GRBs from their host galaxies \citep{Troja+08}; however, the statistical significance of this difference was subsequently challenged \citep{Fong&Berger13}.  

Is such a model consistent with current event rate constraints? A fraction $f_{\rm EE} \gtrsim 0.2-0.4$ of SGRBs are accompanied by extended emission (\citealt{Norris&Bonnell06}). For our progenitor dichotomy scenario to hold, the ratio of the volumetric rate of NS-BH mergers to that NS-NS mergers must be at least as high as $f_{\rm EE}/f_{\rm SGRB}$, where we have assumed that all NS-NS mergers are accompanied by a SGRB but only a fraction $f_{\rm SGRB} < 1$ of NS-BH mergers do the same (as the latter requires rapid BH spin in the prograde direction relative to the orbit for the NS to be tidally disrupted outside the BH horizon).

From the first and second observing runs of Advanced LIGO, the observed rate of NS-NS mergers is $\mathcal{R}_{\rm NSNS} = 110-3840$ Gpc$^{-3}$ yr$^{-1}$ at 90\% confidence, while the upper limit on the rate of mergers of $\gtrsim 5M_{\odot}$ with NSs is $\mathcal{R}_{\rm NSBH} \lesssim$ 610\,Gpc$^{-3}$ yr$^{-1}$ \citep{LIGO+18CATALOG}. The ratio of these rates is thus only weakly constrained at present, such that as long as a moderate fraction of NS-BH mergers produce GRBs,
\be
f_{\rm SGRB} \gtrsim f_{\rm EE}\left(\frac{\mathcal{R}_{\rm NSNS}}{\mathcal{R}_{\rm NSBH}}\right) \gtrsim 0.05\left(\frac{f_{\rm EE}}{0.3}\right),
\ee
one cannot yet rule out the possibility that NS-BH mergers are sufficiently common to account for the population of SGRBs with extended emission ($f_{\rm EE} \gtrsim 0.2-0.4$).  While thus far one NS-NS merger has been discovered compared to zero NS-BH mergers, the statistics of the current sample are obviously very small.  If all extended emission is attributed to fall-back accretion in NS-BH mergers, then our model implies that the steady-state discovery rate of of NS-BH mergers will be significantly higher than that of NS-NS mergers (and that a sizable fraction of the BHs in these systems are spinning in the prograde orbital direction).

\section{Conclusions}

Despite the recent discovery of a short burst of gamma-rays in association with the gravitational waves from a NS-NS merger, the origin of the temporally-extended X-ray emission which is observed following a significant fraction of short GRBs remains a mystery.  Late-time activity from the black hole accretion disk powered by fall-back of marginally bound debris, as would naively be expected in both NS-NS and NS-BH mergers, has been proposed as a source of this behavior (e.g.~\citealt{Rosswog07}).  Following \citet{Metzger+10a}, we have employed a simple model to explore the impact of $\beta-$decay heating due to $r$-process nucleosynthesis on the time-dependence of the mass fall-back rate, using initial data on the ejecta fluid elements (Fig.~\ref{fig:ye_u}), and the properties of the $r$-process heating received (Fig.~\ref{fig:heating_rate}), extracted directly from NS-BH merger simulations.   

We confirm that this $r$-process heating significantly alters the fall-back rate from the canonical $\dot{M} \propto t^{-5/3}$ behavior, generating instead either an abrupt cut-off, or temporal gap, in the fall-back rate on a timescale of $\sim 10-100$ s (Fig.~\ref{fig:fallback}, \ref{fig:param}).  This behavior is robust to the presence of a realistic spread in the heating properties of the fluid elements imparted by a realistic range in the electron fraction and thermodynamic history (Fig.~\ref{fig:Yespread}). Whether a cut-off or gap behavior is obtained depends on the value of a critical dimensionless parameter $\eta$ (eq.~\ref{eq:eta}). The dependence of $\eta$ on black hole mass suggests a possible distinction between cut-off behavior in NS-NS mergers (low black hole mass) and delayed fall-back in NS-BH mergers (high black hole mass), as illustrated in Fig.~\ref{fig:YeM}. The presence or absence, respectively, of extended emission thus provides a possible way to distinguish NS-BH from NS-NS mergers.

Our model could be improved or extended along several fronts in future work. In addition to the BH mass, the critical parameter $\eta$ depends on the total energy $Q_{\rm tot}$ and heating timescale $t_{\rm heat}$ of the $r$-process.  These properties are related to the $Q$ values and $\beta-$decay rates of neutron-rich isotopes whose masses and other properties have yet to be measured by laboratory experiments (e.g.~\citealt{Horowitz+18} and references therein).  A more thorough parameter study of the range of $t_{\rm heat}$ and $Q_{\rm tot}$, e.g. assuming different theoretical models for the nuclear masses, would provide an additional check on the robustness of our conclusions. 

More ambitiously, multidimensional hydrodynamical simulations of the fall-back process, accounting for the effects of $r$-process heating, are needed for a more robust assessment of the gap or cut-off formation process.  Our treatment of directly placing the $r$-process thermal energy into debris kinetic energy neglects the transfer of thermal energy between adjacent fluid elements, though the highly-supersonic and nearly radial motion of the debris should mitigate these effects (Fig.~\ref{fig:schematic}).  Models that do not self-consistently include the back-reaction of the thermodynamics of the fluid elements on the $r$-process path (\citealt{Rosswog+14}) may be inadequate, because once matter starts to recompress and adiabatically heat, the rate of $\beta-$decay heating will be substantially suppressed as the $r$-process path moves back towards the stable valley due to the higher temperatures \citep{Metzger+10a}.  This complex "feed-back"  process may ultimately necessitate coupling at least a simplified $r$-process network (e.g., a one-zone model such as that of \citealt{Lattimer+77}) directly into the hydrodynamical simulations.

\section*{Acknowledgements}

We thank Daniel Kasen for early help on this project.  D.~D.~acknowledges support from the UC Berkeley Summer Undergraduate Research Fellowship. B.~D.~M.~is supported by NASA (grant NNX16AB30G).  We thank Jonas Lippuner and Luke Roberts for providing the SkyNet heating data used in our analysis.  F.~F.~gratefully acknowledges support from NASA through grant 80NSSC18K0565, and from the NSF through grant PHY-1806278.

\label{lastpage}

\end{document}